# Optimization of the pulse width and injection time in a double-pass laser amplifier

**Daewoong Park[1], Jihoon Jeong[1], and Tae Jun Yu[1,2]\***

[1]*Department of Advanced Green Energy and Environment, Handong Global University, Pohang 37554, Korea*

[2]*Global Institute of Laser Technology, Global Green Research and Development Center, Handong Global University, Pohang 37554, Korea*

\**Corresponding author:taejunyu@handong.edu*

We have optimized the input pulse width and injection time to achieve the highest possible output pulse energy in a double-pass laser amplifier. For this purpose, we have extended the modified Frantz-Nodvik equation [1] by simultaneously including both spontaneous emission and pump energy variation. The maximum achieved fluence of the output pulse was 2.4 J/cm$^2$. An input pulse energy of 1 J could be maximally amplified to output pulse energy of 12.17 J, where the optimal values of the pulse width and injection time of the input pulse were 168 $\mu s$ and 10 $\mu s$, respectively, with the effective pump energy being 8.84 J.

# I. INTRODUCTION

Among the various laser parameters, laser pulse width is a crucial factor to be selected carefully, depending on the application [2,3]. In particular, nanosecond and millisecond pulses have contributed significantly to applications in daily lives and practical laser work. For example, nanosecond lasers can be used for UV marking, thin film patterning, and PC board cutting, as their pulse widths are in the moderate regime, offering a balanced throughput and high quality material processing [2]. Microsecond laser pulses have been widely applied for various medical treatments [4–7]. Nd:YAG laser pulses with a width of microseconds and 1064 nm wavelength are useful for treating keloid, hypertrophic, or surgical scars. Moreover, millisecond laser pulses have been used extensively in laser machining, e.g., for cutting, welding, and drilling of machinery materials throughout the fields of automobile, shipbuilding, and aerospace industries, etc. [8–10]. Therefore, as a tool for simulating laser amplifiers, the Frantz-Nodvik (F-N) equation needs to be extended to include these pulse widths and achieve more accurate results. In addition, as the user demands for laser applications expand, designing a laser amplifier that can maximize the output energy while minimizing the related costs is a vital issue. For achieving the latter, we employed a double-pass structure as a laser amplifier. This choice was made based on the following: firstly, in a double-pass amplifier–since the input pulse passes through a single gain medium twice–one gain medium has the same effect as using two. Therefore, it can aid the maximalization of the input pulse energy. Secondly, because of requiring only half the amount of parts compared with amplifiers producing similar output energies, it can achieve a higher level of compactness and cost-effectiveness. Regarding the former: if the input pulse width reaches levels comparable to the fluorescent lifetime of the laser-active ions in the gain medium, the pump energy variation and spontaneous emission must be considered during amplification.

However, there has been no research carried out considering the pump energy variation and spontaneous emission simultaneously with the pulse overlap during the amplification. The F-N equation has been used to calculate the amplified pulse energies for the arbitrary input pulses passing through the laser gain medium [11]. Since the spontaneous emission was neglected in the derivation of the equation, it can be applied only to calculations with this condition satisfied. Neglecting spontaneous emission during the amplification process means that the population of an excited state can return to the ground state only through stimulated emission. For example, in the case of inputting a seed pulse with a duration of 1 $\mu s$ into a Nd:YVO4 amplifier [12], the fluorescent lifetime of the laser-active ions (90 $\mu s$) is significantly longer than the duration of the amplification. Therefore, the pulse amplification ends before the excited ions spontaneously return to their ground state. Thus, in this case, the derivation conditions of the equation are satisfied. In addition, due to the short amplification time, no additionally incoming pump energy is considered during the amplification. Strictly speaking, the existing F-N equation can only be applied if the input pulse width is significantly shorter than the fluorescent lifetime of the laser-active ions.

In this paper, we have extended the F-N equation to consider the spontaneous emission and pump energy variation ignored during conventional simulations of pulse amplification. The suggested method was applied to a flash-lamp-pumped Nd:YAG double-pass amplifier structure, while simultaneously considering the pulse overlap effect as well. We changed the input pulse width and controlled its injection time. We have analyzed how can the output pulse energy be maximized by these parameters through our simulation results. Finally, we have verified that the optimal input pulse width and injection time can be realized within the given conditions to obtain the maximum output pulse energy.

## II. MODIFIED FRANTZ-NODVIK EQUATION

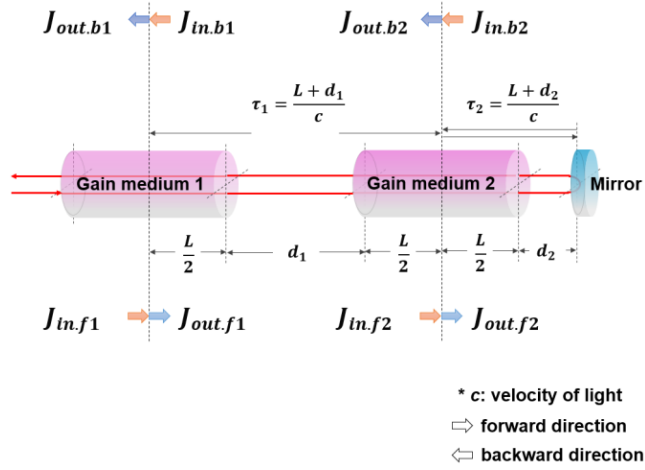

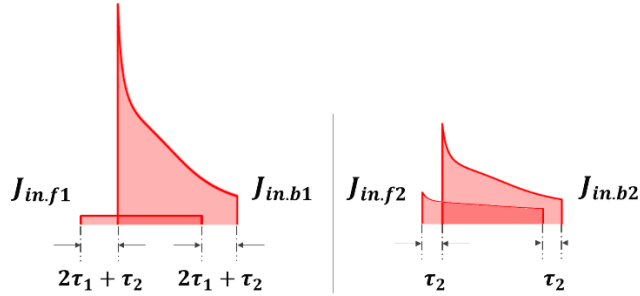

Figure 1. (a) Scheme of a double-pass laser amplifier: $J^{(n)}_{in.f1}$, $J^{(n)}_{out.f1}$, $J^{(n)}_{in.f2}$, and $J^{(n)}_{out.f2}$ are the input and output fluences in the forward propagating direction in the gain media 1 and 2, respectively. Similarly, $J^{(n)}_{in.b1}$, $J^{(n)}_{out.b1}$, $J^{(n)}_{in.b2}$, and $J^{(n)}_{out.b2}$ are the input and output fluences in the backward propagating direction in the gain media 1 and 2, respectively. The temporal lengths $\tau_1$ and $\tau_2$ calculated from the geometry of the amplifier determine the time delay between each input. (b) The front part of the amplified input pulse is reflected on the mirror and is overlapped with the rear part of the input pulse in the gain media 1 and 2, according to $\tau_1$ and $\tau_2$.

Equation (1) gives the modified F-N equation suggested in our previous study which enables the numerical calculation of the pulse amplification in a double-pass amplifier [1]. If the length of an input pulse exceeds the temporal length of the amplifier, as shown in Fig. 1(a), the pulse overlap influences the amplification. Due to this pulse overlap, two inputs and two outputs appear to coexist in the gain medium. The degree of overlap between each input is described in Fig. 1(b). Since there is a common gain medium for both inputs, it can be represented by the simple gain formula $\left(G_E = \frac{J_{out}}{J_{in}}\right)$, as in Eq. (1a). Each output can be obtained by multiplying the respective input by the gain, as shown in Eq. (1b) and Eq. (1c).

$$G_E^{(n)} = \frac{J_{sat}}{J_{in.f}^{(n)} + J_{in.b}^{(n)}} \ln\left\{1 + \left[\exp\left(\frac{J_{in.f}^{(n)} + J_{in.b}^{(n)}}{J_{sat}}\right) - 1\right]\exp\left(\frac{J_{sto}^{(n)}}{J_{sat}}\right)\right\} \qquad (1a)$$

$$J_{out.f}^{(n)} = G_E^{(n)} J_{in.f}^{(n)} \qquad (1b)$$

$$J_{out.b}^{(n)} = G_E^{(n)} J_{in.b}^{(n)} \qquad (1c)$$

$$J_{sto}^{(n)} = J_{sto}^{(n-1)} - \left(G_E^{(n)} - 1\right)\left(J_{in.f}^{(n-1)} + J_{in.b}^{(n-1)}\right) \qquad (1d)$$

$$J_{sto}^{(n)} = J_{sto}^{(n-1)} - \left(G_E^{(n)} - 1\right)\left(J_{in.f}^{(n-1)} + J_{in.b}^{(n-1)}\right) + I_{pump}^{(n-1)} \Delta T - \frac{1}{\tau_f} J_{sto}^{(n-1)} \Delta T \qquad (1e)$$

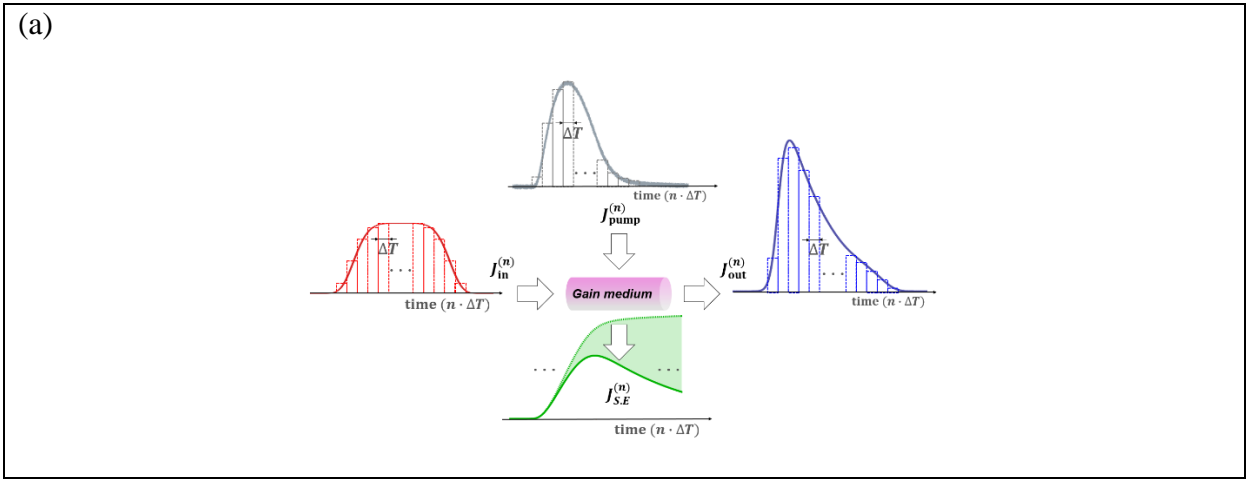

(a)

(b)

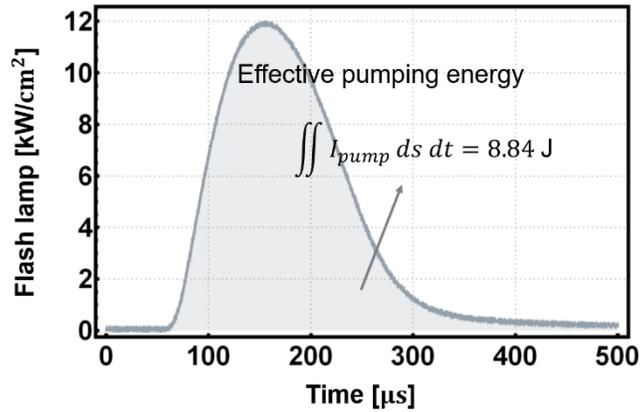

(c)

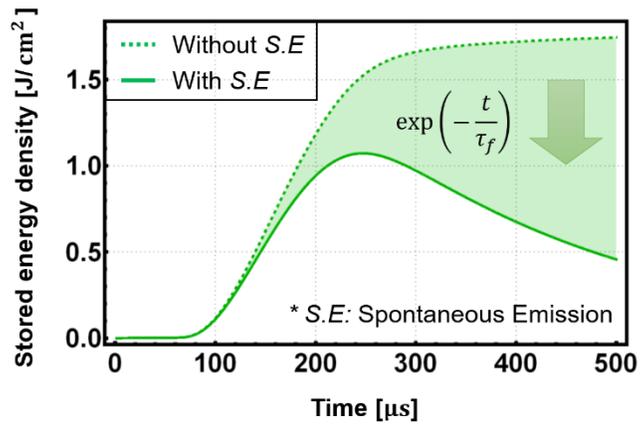

Figure 2. (a) Numerical F-N Equation can be used to calculate the amplification during the whole duration of the input pulse according to the temporal sequence. $J_{in}^{(n)}$, $J_{out}^{(n)}$, $J_{pump}^{(n)}$, and $J_{S.E}^{(n)}$ are the input- and output fluences, effective pump, and spontaneous emission, respectively. (b) Temporal profile of the flash lamp pulse whose effective pump energy is 8.87 J. (c) Variation of the stored energy density in the gain medium pumped by a flash lamp: the spontaneous emission shows an exponential attenuation described by the fluorescent lifetime ($\tau_f$) of the laser-active ions. The solid green line was obtained by considering spontaneous emission, while the green dashed line without.

As shown in Fig. 2(a), the input pulse is divided into rectangular-shaped pulses with discrete lengths of $\Delta T$. The superscript $n$ of each variable is the time index ($t = n \cdot \Delta T$, $n = 0, 1, 2, ..., N$). The amplified fluence of the input pulse can be calculated in a time-sequential manner. Equation (1d) describes the conservation of the stored energy inside the gain medium. The stored energy density $J_{sto}^{(n)}$ of the laser gain medium represents the number density $\Delta_0$ of the excited laser-active ions described in small signal gain, $\exp[\Delta_0 \sigma_e L]$. $\sigma_e$ and $L$ are the emission cross section and length of the gain medium, respectively. If the length of the input pulse is sufficiently short compared with the fluorescent lifetime of the laser-active ions, the stored energy density can be approximated to be influenced only by $J_{in}^{(n)}$ and $J_{out}^{(n)}$ representing the pulse amplification process as terms of stimulated emission. Thus, in Fig. 2(a), the effective pump fluence $J_{pump}^{(n)}$ and the spontaneous emission $J_{S.E}^{(n)}$ can be neglected.

Meanwhile, in the case of pulsed Nd:YAG lasers with a pulse width of microseconds, or even milliseconds, the pulse is not sufficiently short compared with the fluorescent lifetime of the trivalent neodymium ions (230 $\mu s$) used to dope the YAG crystal. The gray curve in Fig. 2(b) gives the shape of the flash lamp pulse used in our lab, measured by an oscilloscope. By assuming that spontaneous emission can be neglected, the laser gain medium would only absorb, without emitting the flash lamp energy. Therefore, the pump energy would be accumulated along the green dashed line in Fig. 2(c). However, as time passes, excited ions fall to an energy level above their ground state through spontaneous emission. Thus, the actual energy accumulated in the gain medium due to flash lamp pump and exponential decay described by the fluorescent lifetime follows the solid green line in Fig. 1(c). Therefore, for the amplification of relatively long input pulses (in the microsecond to millisecond range)–which are comparable to, or longer than the fluorescent lifetime of Nd:YAG–in addition to the input and output fluences, the continuous fluences of the pump source and the spontaneous emission originating from fluoresce must also be considered during the calculation of the stored energy conservation. For the numerical approach, we assume that only the values regarded as constants affect the stored energy. If only these values are considered during a time interval of $\Delta T$, the solution can be presented in a simple form, as given in Eq. (1e).

In Eq. (1e), the effective pump fluence $I_{pump}^{(n-1)} \Delta T$ can be obtained by scaling the pump pulse as the fluence unit $[J/cm^2]$ through the maximum stored energy in the gain medium and the time when it reaches its maximum value. If the input energy is sufficiently low, small signal gain $\exp\left(\frac{J_{sto}^{(n)}}{J_{sat}}\right)$ can be approximated by an overall amplifier gain: $G_E = \frac{J_{out}}{J_{in}}$. Thus, by measuring the small signal gain for a low-energy input, the maximum stored energy $J_{sto.MAX}^{(n)}$ can be obtained. In addition, the time with maximum stored energy can be calculated by the convolution of the pump pulse and exponential decay given by the fluorescence lifetime of the laser-active ions.

The spontaneous emission can be expressed as a partial time derivative of the number density of the excited state, $\frac{\partial}{\partial t} n(t) = -\frac{1}{\tau_f} n(t)$ [13]. It means that the excited ions exhibit an exponential attenuation described by their fluorescent lifetime $\tau_f$. Therefore, the spontaneously emitted fluence can be described by the last term of Eq. (1e).

# III. SIMULATION METHOD AND PARAMETERS

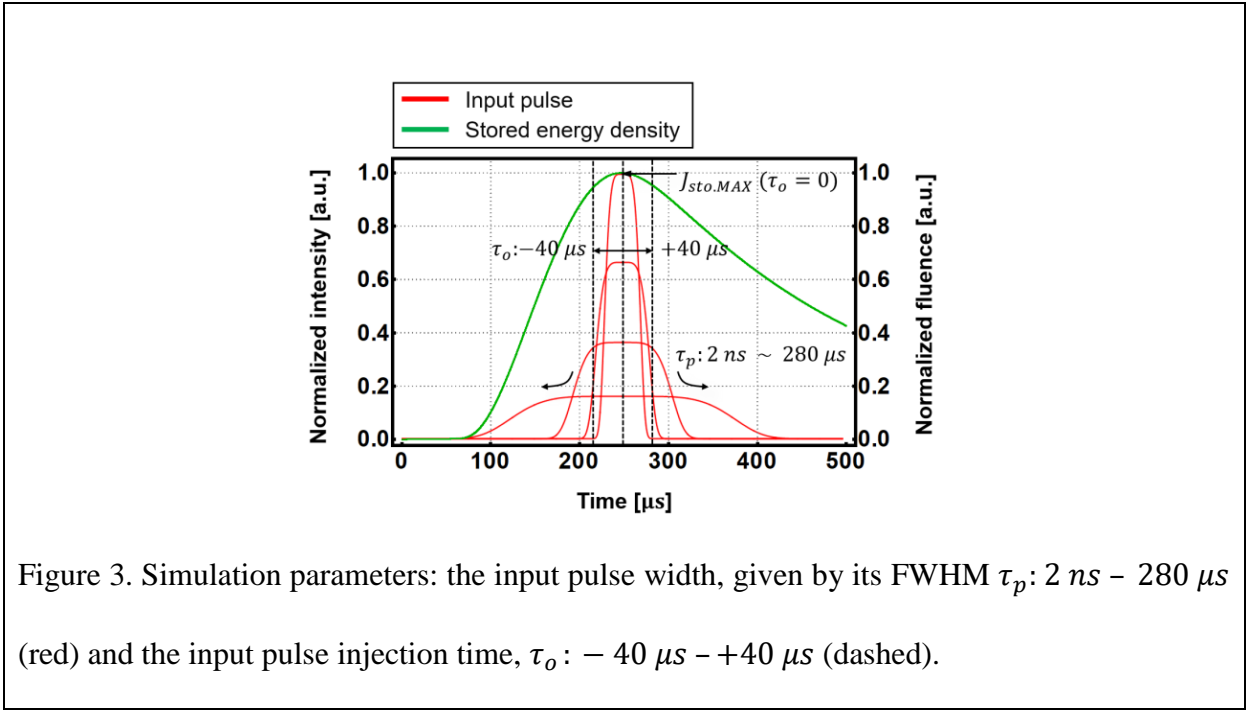

Figure 3. Simulation parameters: the input pulse width, given by its FWHM $\tau_p$: $2\ ns$ – $280\ \mu s$ (red) and the input pulse injection time, $\tau_o$: $-40\ \mu s$ – $+40\ \mu s$ (dashed).

The proposed method was applied to a double-pass amplifier with the aim of maximizing its output pulse energy. In Fig. 1(a), we assumed its geometry as L + $d_1$ = $1500\ mm$, L + $d_2$ = $1500\ mm$. Two Nd:YAG rods having the following properties: $\emptyset = 25.4\ mm$, $\sigma_e = 28 \times 10^{-20}\ cm^2$, $J_{sat} = 0.66\ J/cm^2$ at 1064 nm, $\tau_f = 230\ \mu s$; were used as the laser gain media. The pump source was a flash lamp. We used super-Gaussian pulses of the same order, 5, for the amplifier input. Since the super-Gaussian input pulses have the same energy, 1 J, when the pulse width increases, the peak intensity decreases, as shown in Fig. 3. Thus, the input pulse width affects the shape and energy of the output pulse. The fluence of the output pulse and variation of the stored energy density of each Nd:YAG rod during the amplification were calculated while varying the full-width at half maximum $\tau_p$ and the injection time $\tau_o$ of the input pulse. The injection time was adjusted from $-40\ \mu s$ to $+40\ \mu s$ relative to $\tau_o$ set to zero: we set the injection time to zero when the stored energy in the Nd:YAG rod deposited by the flash lamp reaches its maximum, and it is positioned at the center of the input pulse. We obtained the maximum stored energy, 5.43 J at 247 $\mu s$ after the flash lamp emission. Half of the electrical energy from the power supply was converted into light by the flash lamp, and the efficiency of storing the flash lamp energy in the Nd:YAG was 6.9 %. In addition, the fluence of the output pulse was scanned for input pulse widths from 2 $ns$ to 280 $\mu s$ at each pulse injection time.

## IV. SIMULATION RESULTS AND DISCUSSION

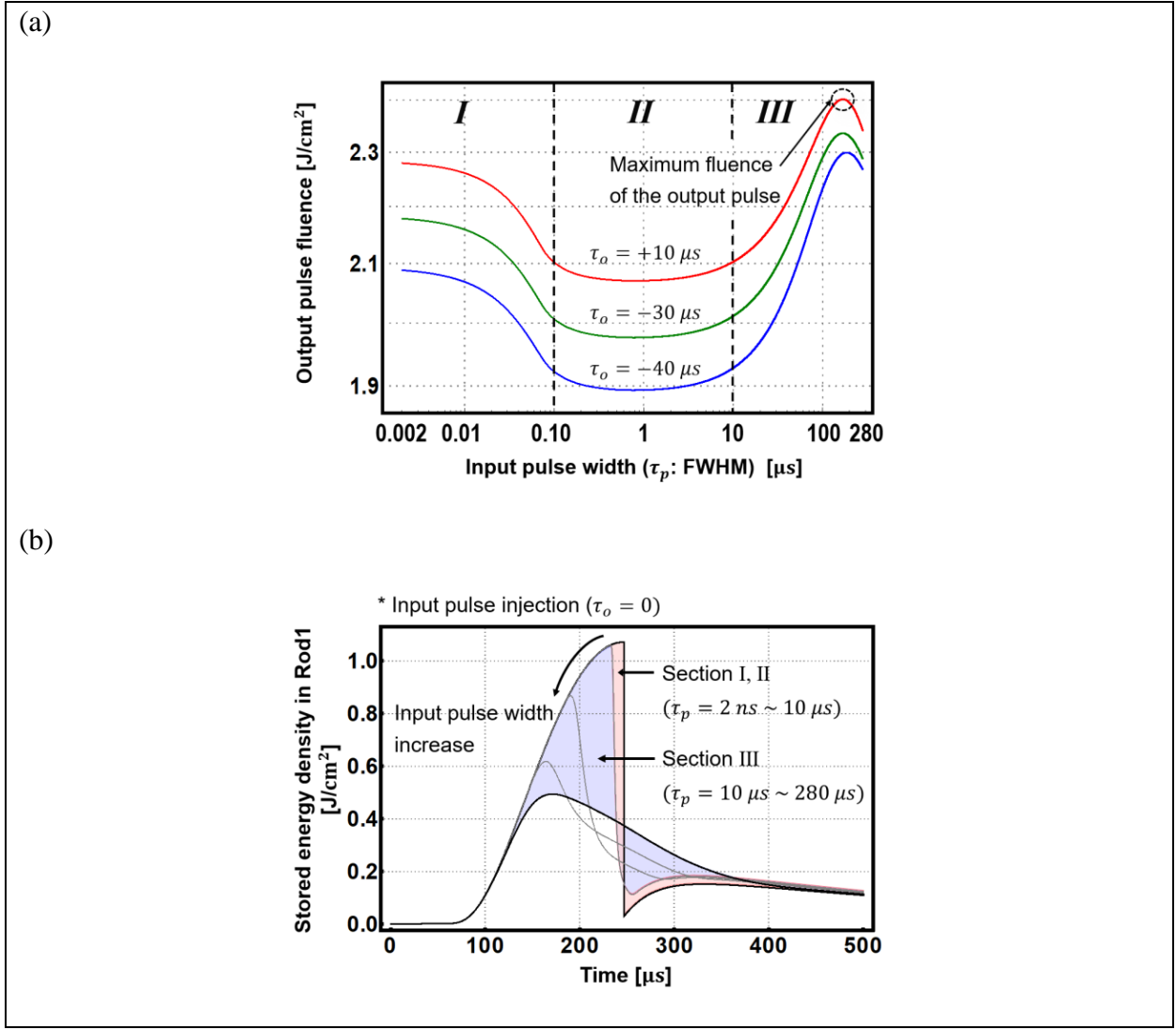

(c)

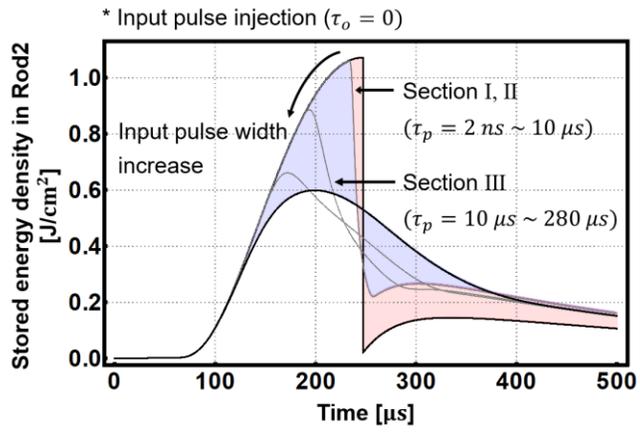

(d)

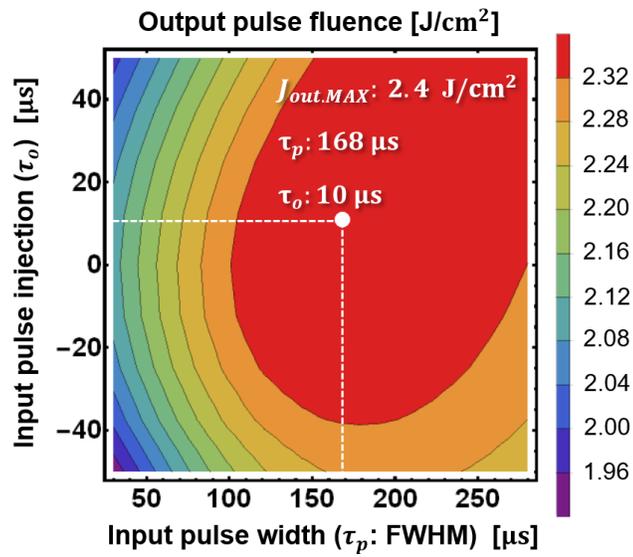

(e)

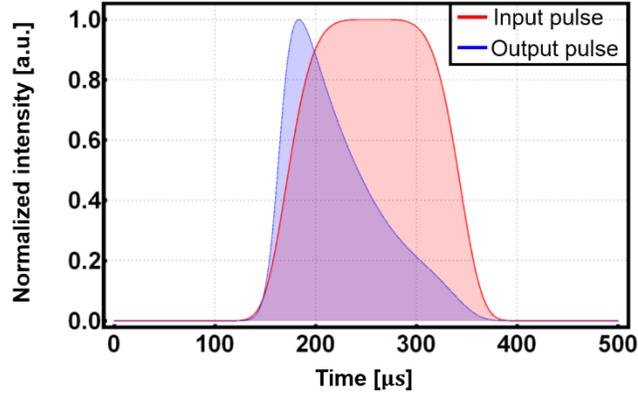

Figure 4. (a) Output pulse fluence as function of the input pulse width where $\tau_o = +10\ \mu s$ (red), $\tau_o = -30\ \mu s$ (green), and $\tau_o = -40\ \mu s$ (blue) (b, c) Trend of the stored energy density in Nd:YAG rod1 and Nd:YAG rod2 during amplification, where $\tau_o = 0$ and $\tau_p = 2\ ns \sim 280\ \mu s$ (d) Optimal pulse width and injection time of the input pulse resulting in the maximum output fluence (e) Normalized input and output pulse shape at the optimal input conditions in (d).

Figure 4(a) shows the fluence variation of the output pulse as function of the full-width at half maximum of the input pulse $\tau_p$. The input pulse width was divided into three sections according to the trend of the fluence variation of the output pulse; section I: $\tau_p \leq 100\ ns$, section II: $100\ ns \leq \tau_p \leq 10\ \mu s$, section III: $10\ \mu s \leq \tau_p \leq 280\ \mu s$. Regarding every injection time of the input pulse ($\tau_o$), the fluence of the output pulse for the given input pulse width tends to show similar variations in all three sections. Therefore, we have considered only the $\tau_o$ values of $+10\ \mu s$ (red), $-30\ \mu s$ (green), and $-40\ \mu s$ (blue).

Firstly, in section I, the input pulse length is remarkably short compared to the fluorescent lifetime of Nd:YAG. Therefore, even if the spontaneous emission and pump energy variation are considered during the pulse amplification, the effect is almost negligible. There is no significant difference from the conventional amplification results. Since the amplification time is concise, the stored energy of the gain medium is mainly influenced by the stimulated emission and is rapidly depleted during the amplification as shown in Fig. 4(b) and Fig. 4(c). To increase the fluence of the output pulse, the input pulse should be injected when the energy stored in the gain medium through pumping is at its maximum. Thus, as shown in Fig. 4(a), the closer to $\tau_o = 0$, the higher the fluence of the output pulse. In addition, as the input pulse width is shortened, the pulse overlap in the double-pass amplifier is reduced, leading to an increased fluence of the output pulse.

When the length of the input pulse belongs to section II, the trend of the output pulse fluence change as function of the input pulse width differs from that of section I. For longer input pulse widths, the fluence of the output pulse in the double-pass amplifier is reduced, as compared with the case of a single-pass amplifier having no overlap. However, as the input pulse width is longer, the pump energy which could not be utilized in the short pulse amplification can be utilized due to extended amplification time. Therefore, this has the effect of increasing the fluence of the output pulse. That is, due to two conflicting effects, the fluence of the output pulse tends to stay almost constant with a varying input pulse width. In addition, the depletion of the excited laser-active ions became more gradual than in section I due to extended amplification time, as shown in Fig. 4(b) and Fig. 4(c). However, this does not make much difference overall, since the length of the input pulse is still considerably shorter than the Nd:YAG fluorescent lifetime. Thus, the effect of spontaneous emission is negligible. To obtain high output pulse energies, it is advantageous to inject the input pulse when the stored energy is as high as possible, similarly to section I.

Finally, in section III, the input pulse width becomes comparable to the fluorescent lifetime of the Nd:YAG. Hence, the pulse overlap effect, the pump energy variation, and the spontaneous emission affect the output pulse fluence in the amplification process. On the left-hand side of section III in Fig. 4(a), the fluence of the output pulse tends to increase as the input pulse width becomes gradually longer. In this case, the effect of using more pump energy by longer amplification time has a more dominant influence on the output fluence than the energy loss caused by pulse overlap and spontaneous emission. On the contrary, the right-hand side of section III in Fig. 4(a) shows an opposite trend. Thus, the output fluence reaches its maximum value and then subsequently decreases. As a result, in section III, we could find the optimal injection time and input pulse width, with which the maximum output fluence is achieved. From the considered simulation parameters and conditions, the maximum achievable output fluence and energy are 2.4 $J/cm^2$ and 12.17 J, respectively, with an optimal injection time $\tau_o$ of $+10\ \mu s$; optimal full-width at half maximum $\tau_p$ of 168 $\mu s$; input pulse energy of 1 J; and effective pump energy of 8.84 J.

## IV. CONCLUSION

We have studied the design of a laser amplifier that can maximize its output pulse energy. For this purpose, we have chosen a double-pass amplifier structure that can utilize one laser gain medium twice, leading to economic benefits of its manufacturing and a compact design. Secondly, a pulse amplification simulation method, considering spontaneous emission and pump energy variation numerically, has been proposed by extending the F-N equation. The original F-N equation is limited for certain input pulse width range by the assumptions made for its derivation, e.g., neglecting spontaneous emission. The proposed method allows a more accurate amplification simulation of pulses with widths comparable or longer than the fluorescent lifetime of the laser-active ions. The method is applicable for amplifiers other than the demonstrated Nd:YAG double-pass structure as well. Finally, we set the simulation parameters to the super-Gaussian input pulse width and the injection time, while considering the pulse overlap effect during the amplification. We observed the output pulse energy and the stored energy variation in the gain medium to obtain the highest achievable output pulse energy. With the considered parameters, for pulse widths below 100 $ns$, i.e., for pulse widths significantly shorter than the fluorescence lifetime of the laser-active ions, the result fulfilled our intuitive expectations. Short pulses injected when the stored energy of the gain medium is at its maximum can produce the highest possible output pulse energy through abrupt depletion of the upper-state ions. This is caused by the shorter pulses reducing the energy wasted by spontaneous emission and temporal overlap. In addition, as an intermediate region, input pulse widths between 100 $ns$ and 10 $\mu s$ do not exhibit a noteworthy influence on the output pulse energy. Finally, beyond 10 $\mu s$ pulse widths, we obtained the optimal input pulse width and injection time resulting in the maximum output pulse energy. Consequently, this paper is expected to be useful for designing pulsed laser amplifiers with the aim of generating more energy under the same conditions.

## ACKNOWLEDGMENT

This work was supported by the Industrial Strategic Technology development program, 10048964. The development of the 125 J·Hz laser system for laser peening was funded by the Ministry of Trade, Industry & Energy (MI, Korea).

# FIGURE (PHOTO) DESCRIPTION

Figure 1. (a) Scheme of a double-pass laser amplifier: $J_{in.f1}^{(n)}$, $J_{out.f1}^{(n)}$, $J_{in.f2}^{(n)}$, and $J_{out.f2}^{(n)}$ are the input and output fluences in the forward propagating direction in the gain media 1 and 2, respectively. Similarly, $J_{in.b1}^{(n)}$, $J_{out.b1}^{(n)}$, $J_{in.b2}^{(n)}$, and $J_{out.b2}^{(n)}$ are the input and output fluences in the backward propagating direction in the gain media 1 and 2, respectively. The temporal lengths $\tau_1$ and $\tau_2$ calculated from the geometry of the amplifier determine the time delay between each input. (b) The front part of the amplified input pulse is reflected on the mirror and is overlapped with the rear part of the input pulse in the gain media 1 and 2, according to $\tau_1$ and $\tau_2$.

Figure 2. (a) Numerical F-N Equation can be used to calculate the amplification during the whole duration of the input pulse according to the temporal sequence. $J_{in}^{(n)}$, $J_{out}^{(n)}$, $J_{pump}^{(n)}$, and $J_{S.E}^{(n)}$ are the input- and output fluences, effective pump, and spontaneous emission, respectively. (b) Temporal profile of the flash lamp pulse whose effective pump energy is 8.87 J. (c) Variation of the stored energy density in the gain medium pumped by a flash lamp: the spontaneous emission shows an exponential attenuation described by the fluorescent lifetime ($\tau_f$) of the laser-active ions. The solid green line was obtained by considering spontaneous emission, while the green dashed line without.

Figure 3. Simulation parameters: the input pulse width, given by its FWHM $\tau_p$: $2\ ns - 280\ \mu s$ (red) and the input pulse injection time, $\tau_o$: $-40\ \mu s - +40\ \mu s$ (dashed).

Figure 4. (a) Output pulse fluence as function of the input pulse width where $\tau_o = +10\ \mu s$ (red), $\tau_o = -30\ \mu s$ (green), and $\tau_o = -40\ \mu s$ (blue) (b, c) Trend of the stored energy density in Nd:YAG rod1 and Nd:YAG rod2 during amplification, where $\tau_o = 0$ and $\tau_p = 2\ ns \sim 280\ \mu s$ (d) Optimal pulse width and injection time of the input pulse resulting in the maximum output fluence (e) Normalized input and output pulse shape at the optimal input conditions in (d).

## FIGURE (PHOTO)

Figure 1

(a)

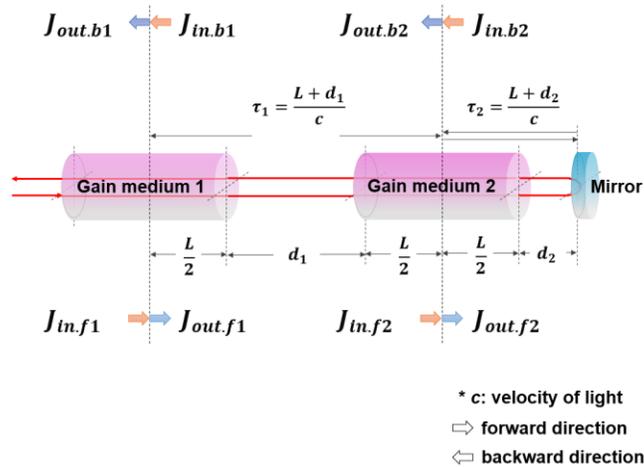

(b)

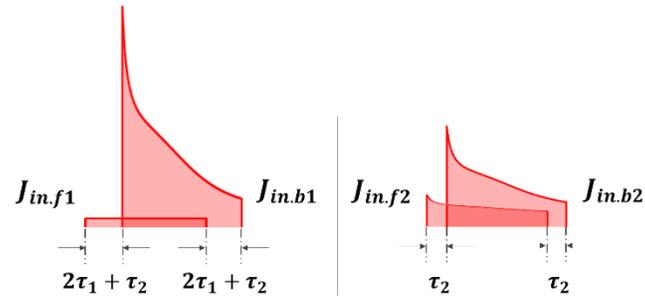

Figure 2

(a)

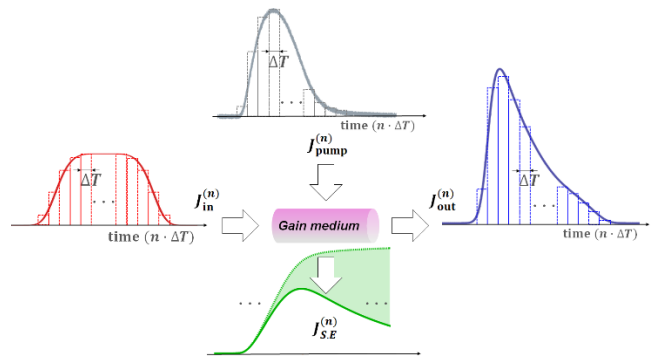

(b)

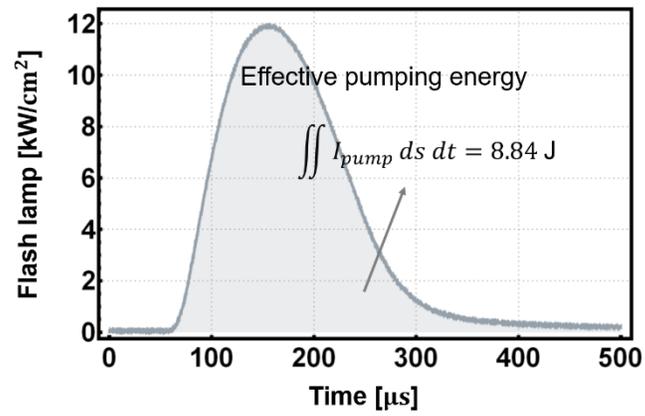

(c)

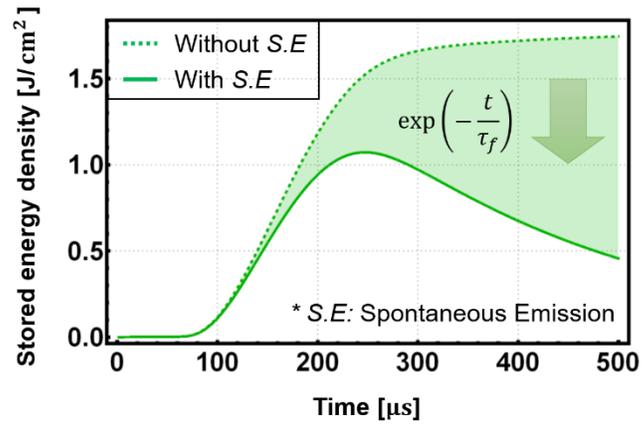

Figure 3

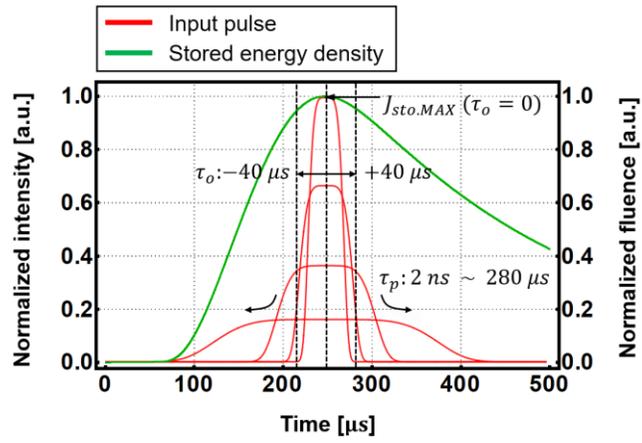

Figure 4

(a)

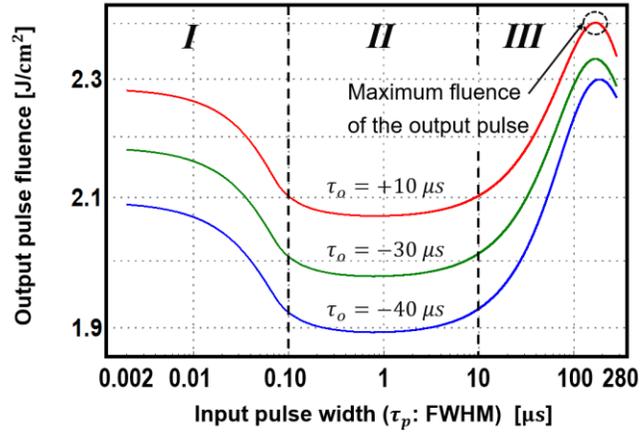

(b)

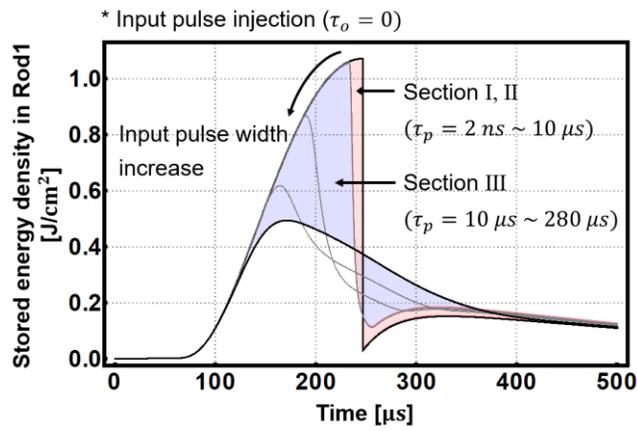

(c)

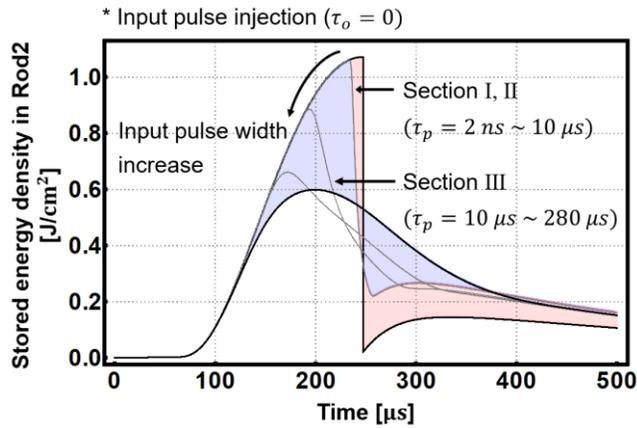

(d)

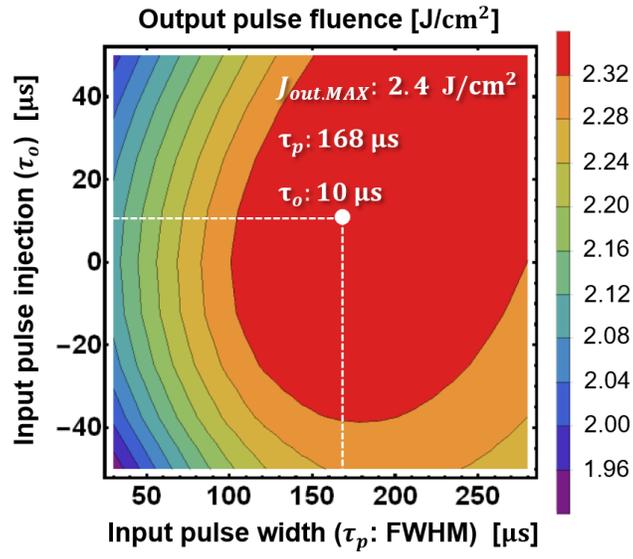

(e)

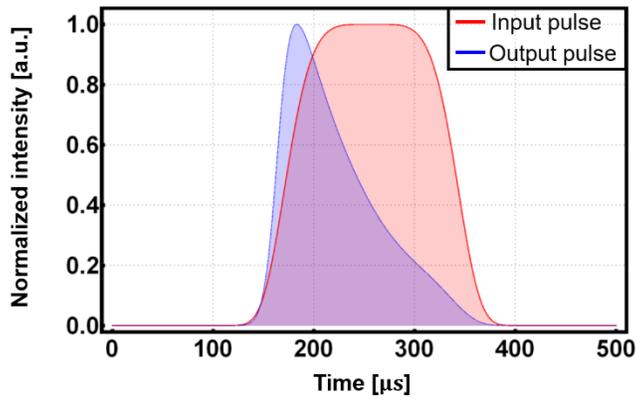